# Neural Mechanisms of Human Decision-Making

Seth Herd[1], Kai Krueger[1,2], Ananta Nair[1], Jessica Mollick[1,3], and Randall O'Reilly[1,4]

[1] eCortex, Inc.  [2] University of Colorado, Boulder  [3] Yale University, New Haven, CT  [4] University of California, Davis

**Abstract**:
**We present a computational and theoretical model of the neural mechanisms underlying human decision-making. We propose a detailed model of the interaction between brain regions, under a *proposer-predictor-actor-critic* framework. Task-relevant areas of cortex *propose* a candidate plan using fast, model-free, parallel constraint-satisfaction computations. Other areas of cortex and medial temporal lobe can then *predict* likely outcomes of that plan in this situation. This step is optional. This prediction-(or model-)based computation produces better accuracy and generalization, at the expense of speed. Next, linked regions of basal ganglia *act* to accept or reject the proposed plan based on its reward history in similar contexts. Finally the reward-prediction system acts as a *critic* to determine the value of the outcome relative to expectations, and produce dopamine as a training signal for cortex and basal ganglia. This model gains many constraints from the hypothesis that the mechanisms of complex human decision-making are closely analogous to those that have been empirically studied in detail for animal action-selection. We argue that by operating sequentially and hierarchically, these same mechanisms are responsible for the most complex human plans and decisions. Finally, we use the computational model to generate novel hypotheses on causes of human risky decision-making, and compare this to other theories of human decision-making.**

Introduction

Decision-making is of critical importance. In personal life, professional activities, and in government and military contexts, the quality of people's decisions is among the most important determinants of whether our outcomes are good, bad, or disastrous. As such, a great deal of scientific work has been directed at human decision-making, at multiple levels of analysis. In this paper, we advance an integrated computational and theoretical framework for understanding how specific brain networks give rise to both the power and pitfalls of human-level decision making, building upon a foundation of existing functional and anatomical studies in animals and behavioral and neuroimaging studies in humans.

There is now a broad consensus about the critical role of the basal ganglia in helping to select actions. By learning over time from dopamine neuromodulation, it selects those actions which maximize reward and minimize negative outcomes (Barto, Sutton, & Anderson, 1983; Barto 1995; Mink, 1996; Graybiel, Aosaki, Flaherty, & Kimura, 1994; Joel, Niv, & Ruppin, 2002; Graybiel, 2005; Nelson & Kreitzer, 2014; Gurney, Prescott, & Redgrave, 2001; Frank, Loughry & O'Reilly, 2001; Brown, Bullock, & Grossberg, 2004; Frank, 2005; O'Reilly & Frank, 2006; see Collins and Frank, 2014 and Dunovan & Verstynen, 2016 for recent reviews). Computationally, this system is well-described by the *Actor-Critic* framework of Sutton and Barto (1981). In this framework, the basal ganglia action-selection system is the "actor", and the set of brain areas that produce phasic changes in dopamine are termed the "critic". This system evaluates selected actions and improves choices over time, while also improving the critic itself.

More recently, there has been considerable interest in a higher-level, *model-based* form of action selection, thought to depend on prefrontal cortical areas. This has been contrasted to the *model-free* nature of the learned associations in the basal ganglia system (Daw, Niv, & Dayan, 2005; Dayan and Berridge, 2014).Many people tend to think of this distinction in terms of separate, and perhaps competing, systems that enact *goal-directed* versus *habitual* behavior (Tolman, 1948; Balleine & Dickenson, 1998; Yin & Knowlton, 2006; Tricomi, Balleine & O'Doherty, 2009), where the basal ganglia is the habit system, and the prefrontal cortex is goal-directed. However, it is increasingly clear that the basal ganglia plays a critical role in higher-level cognitive function (e.g., Pasupathy & Miller, 2005; Balleine, Delgado, & Hikosaka, 2007), and in goal-directed behavior (Yin, Ostlund, Knowlton, & Balleine, 2005).

We present an alternative model in which the basal ganglia and cortex are not separate, and do not compete, but interact to produce a spectrum of computations. These range between fully model-free (or *prediction-free*), to fully model-based (or *prediction-based*). We use the terms *prediction-based* and *prediction-free,* to avoid a variety of accumulated terminological baggage (see Discussion section and O'Reilly, Nair, Russin and Herd, submitted).

In the context of these existing ideas and issues, we offer a specific theory of how brain systems computationally perform complex human decision-making. In this biologically-based *Proposer-Predictor-Actor-Critic* framework, the prefrontal cortex and basal ganglia work together as an integrated system, and prediction-based computations are an optional additional step, rather than a separate system. In addition, this framework provides an explicit attempt to account for the temporally-extended, sequential nature of complex human-level decision-making. Most existing biologically-constrained theories have been focused on a parallel evaluation of multiple alternatives in a single step, which we argue is only plausible in well-learned tasks (including most laboratory studies). We argue that making important decisions in complex, novel situations demands a slower, serial computational approach to maximize accuracy, flexibility, and transfer of prior learning

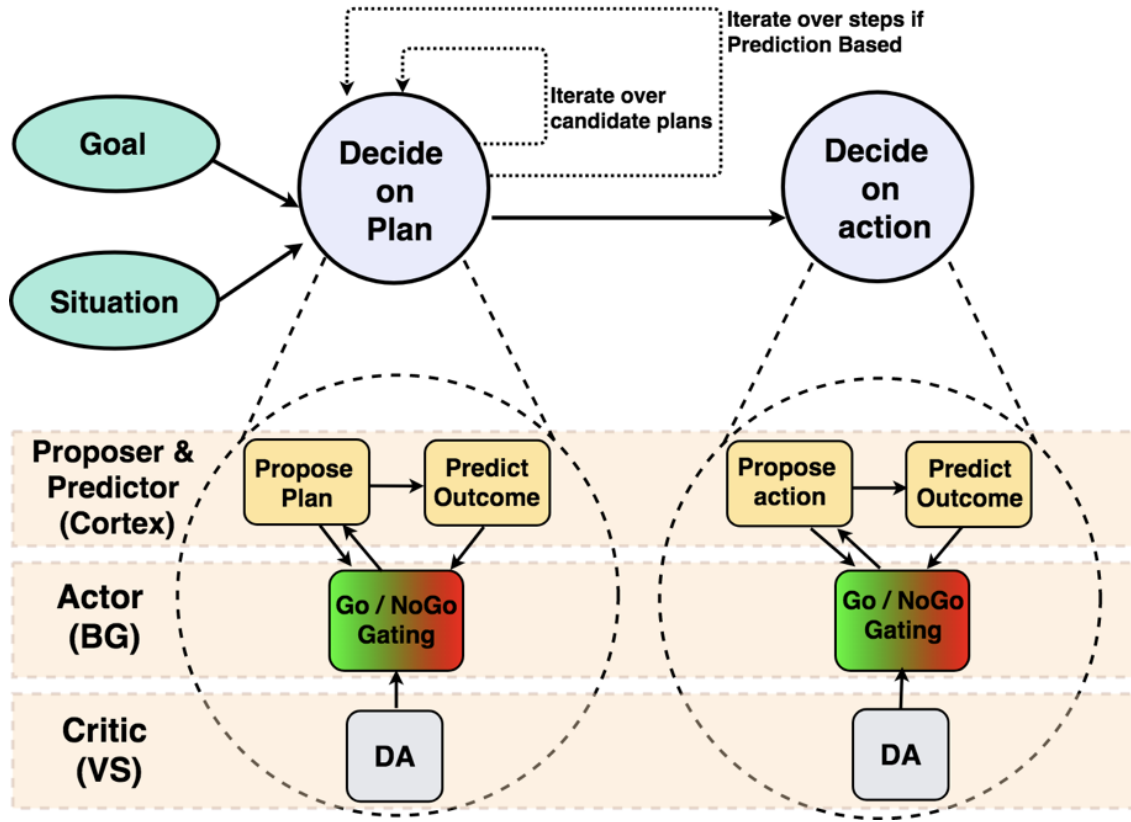

**Figure 1**: Structure of Proposer-Predictor-Actor-Critic architecture across frontal cortex and subcortical areas. We depict two parallel circuits with a hierarchical relationship. The top is a broad functional diagram, emphasizing the serially iterative and hierarchical nature of our proposed decision-making process. The bottom expands those functions, and identifies the brain areas that perform each function. This architecture is described in more detail in the Methods section.

The *Proposer-Predictor-Actor-Critic* circuit (Figure 1) provides a functional model for the prototypical loops descending from all areas of frontal cortex through the basal ganglia and converging back to modulate the function of matching areas of frontal cortex (Alexander, De Long, & Strick, 1986; Haber, 2010; Sallet et al, 2013; Haber, 2017). This same functional circuit operates in each of the different levels of decision-making and action-selection associated with each different fronto-striatal area. We assume, with many others (e.g., Miller & Cohen, 2000) that complex decision-making consists of selecting working memory representations for maintenance, and allowing those representations to condition further steps in the decision-making process. Critically, we also argue that each such circuit also functions sequentially across multiple iterations within any sufficiently complex decision-making task. This is a *serial-parallel* model, in which parallel computations are iterated serially.

- The cortical *Proposer* (e.g., premotor cortex for motor actions) settles on a representation of one potentially rewarding action, plan, or task set (we use the term *Plan* for all of these, since the neural mechanisms are isomorphic). The parallel process of generating a candidate Plan involves bidirectional activation flow across multiple interconnected cortical areas, with iterative *constraint satisfaction* processing that integrates learned synaptic weights with the current external inputs (stimuli, context, affective and body states, etc) to produce a plausible plan that represents at least a locally-maximal satisfaction of all these factors (Hopfield, 1984; Ackley, Hinton & Sejnowski, 1985; Rumelhart & McClelland, 1986; O'Reilly & McClelland, 2000; O'Reilly, Wyatte, Herd, 2013). This process acts in a single parallel neural computation, and so is relatively fast.
- A cortical *Predictor* (e.g., parietal and/or medial temporal lobe for motor actions) predicts *specific outcomes* of that plan in a given context or situation (in terms of outcome states). This step is optional; it is engaged only when a prediction-based strategy is employed, which takes extra time, while (usually) providing additional accuracy and generalization to the decision. Indeed, this decision of whether to engage additional predictions is one of many sequential steps in the overall decision making process.
- The *Actor*, consisting of a basal ganglia loop linked to the *Predictor* area, takes that prediction as input, and uses what it has learned about the linked reward history to accept or reject that plan. If the proposed plan is rejected, the Proposer proposes a different plan, and the process continues. The critical computational feature of this basal ganglia system, which is not well-supported in the cortical system, is the ability to boil everything down to two opposing evaluations: Go (direct) and NoGo (indirect). These pathways compete directly to produce an overall net evaluation that is directly informed by a history of learning via phasic dopamine signals from the Critic. This is the core principle of the widely-accepted animal action-selection model of basal ganglia function. We think it often operates in a more generalized, sequentialized fashion in human decision-making. The basal ganglia receives more highly-processed, abstract information from cortical inputs, which enables it to function effectively in novel decision-making contexts where the history of learning is only applicable when it is associated through abstract representations developed by cortex. One critical feature of this system is its computational power to stop an action, even when that action is already abundantly represented in the cortex. We think this stopping is the basis of humans' ability to systematically consider multiple possible plans or actions before executing one.
- Once a plan is selected and an outcome is experienced, the *Critic* estimates the value of that outcome relative to its expectations for that situation as a *reward prediction error* (Schultz, 2016). This Critic is composed of a set of subcortical areas that function as a reward-prediction system, and it uses that reward prediction to discount the outcome's value, sending the result as a phasic dopamine signal. The Critic's dopamine signal trains both the Actor and the Proposer components, while the Predictor is trained by the specific outcome that occurred.

By binarizing (accept/reject), as well as sequentializing (considering one proposed plan at a time), this canonical decision circuit could scale out to arbitrarily complex decisions, in the same way that sequential computer programs can accommodate arbitrarily complex chains of logic, whereas parallel algorithms have more constraints. For example, the decision of whether to employ prediction, as noted above, is one sub-decision, and each step of modeling the decision tree is another. Furthermore, interactions between isomorphic loops at different levels enable this set of mechanisms to function semi-hierarchically, where higher-level decisions can be unpacked into sub-goals and steps at lower levels.

**Neural Evidence of Sequential Decision-Making**

While most of the neuroscience data in animal models is consistent with the idea that multiple options are evaluated in parallel (e.g., Balleine, Delgado, & Hikosaka, 2007; Collins & Frank,

2014), there is some recent detailed neural recording data consistent with a more serial process. Hunt et al (2018) concluded that monkey orbitofrontal cortex (OFC) and anterior cingulate cortex (ACC) neurons represent the value of the currently attended stimulus. They used a two-alternative choice task, and recorded from monkey frontal neurons. The activation values of those neurons in OFC correlated with the identity and the value of the currently attended stimulus, and the ACC showed a more stepwise function, perhaps functioning as a belief updating and accept/reject signal in their paradigm. OFC primarily represents the value of the currently fixated option; while previous cue values are represented above baseline level, the current cue representation is actually anticorrelated with previous values. This indicates a comparison-with-current-best-option representation, as opposed to the value-summation representations assumed by parallel models. The primary OFC representation data is shown in Figure 2a, and we address other aspects of their data in the Discussion section.

Rich & Wallis (2016) also examined firing patterns in OFC neurons, and found that they largely represented the value of a single choice option at any given point in time. Activity was recorded in OFC while monkeys performed a two-alternative-forced-choice task, and that data was used to train a decoder to distinguish neural representations of those choices. The decoder's estimate of the presence of each option showed a strong reciprocal relationship; a strong representation of one option corresponded to a weak representation of the other option. This pattern of results indicates that the OFC primarily represented the value of one of the two options at any given time point. This data is also summarized in Figure 2(b).

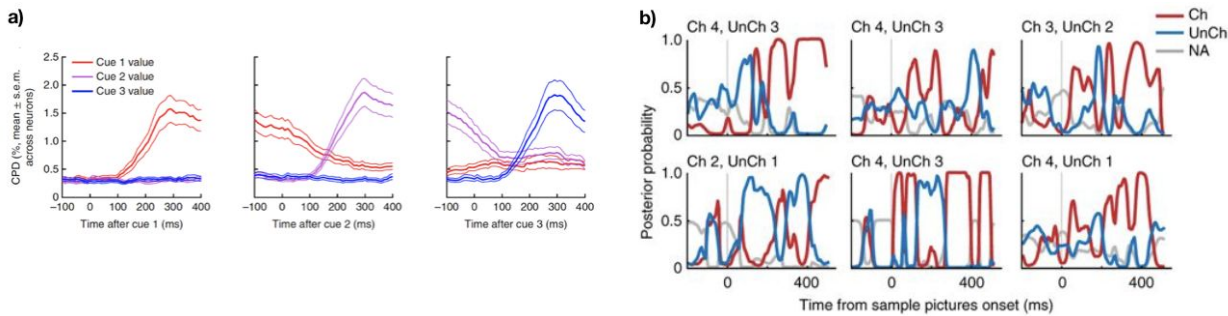

**Figure 2**: **a)** Data from Hunt et al, 2018, showing correlations between neural activity and cue values at each time point in monkey OFC during a two-option comparison task. **b)** Data from Rich & Wallis (2016). Decoder results on individual trials. Red lines are decoder estimates of the probability of a representation of the chosen option; blue lines are estimated probabilities of available of unchosen options; and gray lines are an average of the two remaining options that are not available on that particular trial. Results, and additional analysis, strongly indicate that OFC neurons alternate between representing each of the available options, and do not represent them in parallel.

Despite the overall serial nature of the full loop of decision-making in our framework, there is still an important parallel weighing of options by the Proposer's constraint-satisfaction settling mechanism, to choose an initial plan to evaluate. This is consistent with the finding that value representations in ventromedial prefrontal cortex and striatum in the early stages of decision making tasks correlate strongly with the outcome the animal will go on to select on that individual trial, rather than an average of all available outcomes, as shown by the recording data above, and in Kable & Glimcher (2009).

We ran a full model, as described above, and several variants that illustrate different possible decision-making strategies and brain computations accompanying them, and individual differences in risk bias. Those simulations and results are as follows:

- The full model. Its performance speeds up as it increasingly relies on the prediction-free Proposer component and requires the serial consideration of fewer plans by the Predictor, Actor, and Critic components. This illustrates a smooth transition from more *controlled* to more *automatic* processing (Shiffrin & Schneider 1977) or habitual behavior (Tolman, 1948).
- A no-Proposer variant, to illustrate the role of the Proposer in the full model. Plans are selected for consideration at random instead of through reward learning. This variant performs more slowly, but generalizes to the held-out test set better, since it cannot transition to the more automatic, habitual mode of performance with increasing experience, as the full model does.
- A no-Predictor, model-free variant, in which the model's Proposer and Actor components (cortex and basal ganglia) perform model-free RL. This variant addresses the possibility that the system sometimes makes decisions without taking time to make any prediction about outcomes. This variant performs poorly on our full, complex task, but can perform well on simpler tasks, and performs faster without the need to wait for an explicit prediction form the cortical Predictor layers. We think this is the fastest but least accurate mode of human decision-making, as proposed by Daw & Dayan (2005).
- A value-only Predictor variant, in which the two cortical prediction layers do not predict a specific Result or Outcome, but only the value of the result. This variant blurs the line between prediction-based and prediction-free strategies; it is technically performing a model-free computation, but it is using prediction. This fast, mixed mode of prediction can be performed in a single computational step, and so performs faster but with poor generalization relative to the full model with its two-step Predictor component.
- Finally, we more specifically addressed risky decision-making. We shifted the balance between Go (D1) vs NoGo (D2) pathways in the Actor (basal ganglia) component. The model reproduces experimental results showing more risky decisions with more D1 influence, in accord with a variety of empirical results. We also modeled vicarious learning, in which people learn from others' experiences in risky domains, and showed how different vicarious experiences (E.G., different public awareness campaigns), produce different behavioral risk profiles.

Next, we present the specific computational implementation of our overall framework, the above manipulations, and their results. In the Discussion, we consider the relationships between this framework and a variety of other approaches to understanding decision making across a range of different levels of analysis.

## Methods:

**Task:**

Our model primarily performs an abstract "plan selection" task. We used an abstract task structure in order to address the essential features of decision-making across a range of circumstances. Similar task structures are found in a variety of domains, including social decision-making, and complex task planning used for planning tasks, projects, or errands, etc. While different areas of cortex and basal ganglia perform different types of decision-making, a core hypothesis is that their circuits and computations are closely analogous, so this theory applies to all of those types of decisions.

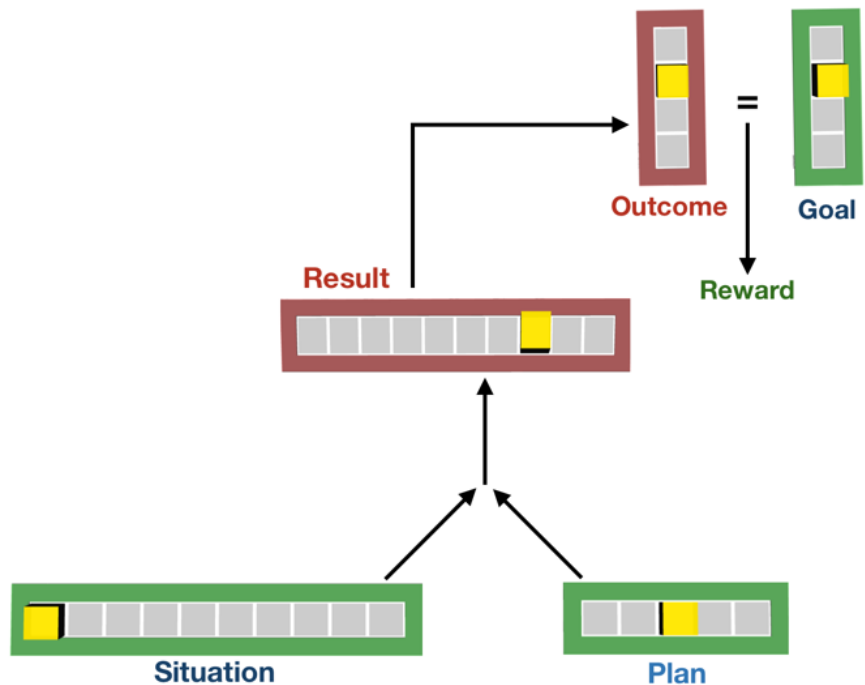

**Figure 3: Task Diagram.** The depicted are input and output layers of the model. The model's task is to choose a Plan that achieves an Outcome matching its current Goal, given a random starting Situation. Each Situation and Plan lead deterministically to a Result (conceptually, another Situation), which leads to one Outcome which is potentially rewarding (e.g., food, water, or shelter, or in a social context, agreement, submission, or appreciation). The model receives a reward signal if it chooses a Plan that matches its current, randomly chosen Goal. There are ten Situations and Results, five Plans, and four Outcomes and Goals, for a total of 240 Plan-Situation-Goal combinations, each of which deterministically leads to success or failure.

In this task, there are Situations, Plans, Outcomes, and Goals. The model is presented with a Situation and a Goal. It iteratively chooses a plan to evaluate, and either accepts or rejects that Plan. When the model selects a Plan, it experiences the Outcome and learns from it. When it rejects a Plan, the Proposer component chooses a different Plan for the Predictor to evaluate. It continues this process until it has either accepted a plan, or run out of time to make a decision.

This task structure applies to many types of decision-making. In spatial foraging, physical locations would be the Situations and Results, Plans would be for navigating between locations, and Outcomes and Goals would be physical requirements like food, water, and shelter. In social decision-making, Situations and Results would be social situations such as meeting a new colleague, being challenged on one's claims, etc, while Plans would be general approaches to social interaction (agree, discuss, threaten, etc), and Goals/Outcomes would be social goals such as gaining respect, gaining sympathy, getting agreement, etc. In complex tasks, Situations and Results would be task states, while Outcomes and Goals would be progress toward different sub-goals or metrics of task completion. For instance, a Situation could be a certain board position in chess, while a Result would be a new board position, which could accomplish

Goals of controlling the center of the board, freeing pieces, creating attacking pressure, or defending a vulnerable piece.

Each Plan and Situation pair lead deterministically to a Result, and each Result always produces the same one Outcome. If that Outcome matches the current Goal, a reward is provided to the model; if it does not, no reward or punishment is given (except for use of stochastic punishment only for the risky decision-making model manipulations, described in that section). In either case, only once a Plan is accepted does the model experience and learn from the outcome. After a plan is accepted, or five plans are rejected in a row (i.e, the model "gives up), training moves on to a new combination of Situation and Goal.

The main results we report here use the following environment and reward structure: There are ten Situations, five Plans, and four Outcomes, matched by four Goals. There are also ten Results, each of which lead deterministically to one of the four Outcomes.

Based on this information, the model can learn the underlying environmental contingencies. Importantly, the cortical, predictive portions of the model learn what Result and Outcome result from the given Plan and Situation combination. The Proposer and the basal ganglia learn based on the reward signal, so they learn to respectively select and accept a plan that produces an Outcome matching the current Goal in the current Situation.

We also used a holdout test set to test generalization. During training, a subset of Situation/Goal pairs were withheld and never shown. This allowed us to test generalization or transfer to novel combinations of situations and goals. In testing mode, those withheld pairings were presented and no weight updates were performed. Testing for 5 epochs is interleaved every 25 epochs during all simulations, to obtain a learning curve for generalization performance for the held out examples. Out of the possible 40 Situation/Goal pairs, for any given simulation, 4 pairs were left out. For each simulation, different Plan/Situation to Result to Outcome pairings were chosen at random, and a different training/testing split was chosen.

**Model:**

Our model was created within the Leabra modeling framework (O'Reilly & Munakata, 2000; O'Reilly et al, online textbook; O'Reilly, Herd, Hazy 2016). The Leabra framework is a cumulative theory of cortical function, with variants covering subcortical function. It has been used to model a wide variety of cognitive phenomena, and is an attempt to constrain a general theory of cortical function with as much evidence as possible. In this case, however, we have focused not on the specific contributions of the Leabra framework, but the general computational properties. Our results hold true for a variety of parameter choices within the Leabra algorithm, and we believe that they should hold true for neural network models with similar architectures, under a wide variety of learning rules, activation functions, and other parameter choices.

The Leabra framework uses point neurons with sigmoidal response functions. It is here (and most frequently in other work) run with a rate-coded responses. This arrangement is similar to a number of other modeling frameworks designed to address similar levels of analysis, those reaching up to human cognition and behavior (Deco & Rolls 2003; Deco & Rolls 2002; Rumelhart & McClelland, 1986; Grossberg 2013, Brown, Bullock & Grossberg 2004; Collins and

Frank, 2014). Leabra units are usually considered to be representative neurons among a much larger population. The model here is of limited size; it contains a total of 975 units, and each of the three cortical processing layers (Proposer, State Predictor and Outcome Predictor) contain only one hundred units each for the standard model. This small model is adequate to demonstrate our general points; much larger models are needed to process complex real-world input (e.g., visual object recognition from images; O'Reilly, Wyatte, Herd, Mingus, & Jilk, 2013).

Specifically, there is one input unit for the four possible current Goals; ten input units for the Situations; ten units for the Result that act as outputs for the State Predictor layer, and inputs to the Outcome Predictor layer; four units for the four possible Outcomes that are outputs from the Outcome Predictor layer; and a Value, a *scalar value* layer which represents the predicted value of the predicted Outcome given the current Goal on a continuous -1 to 1 punishment to reward scale.

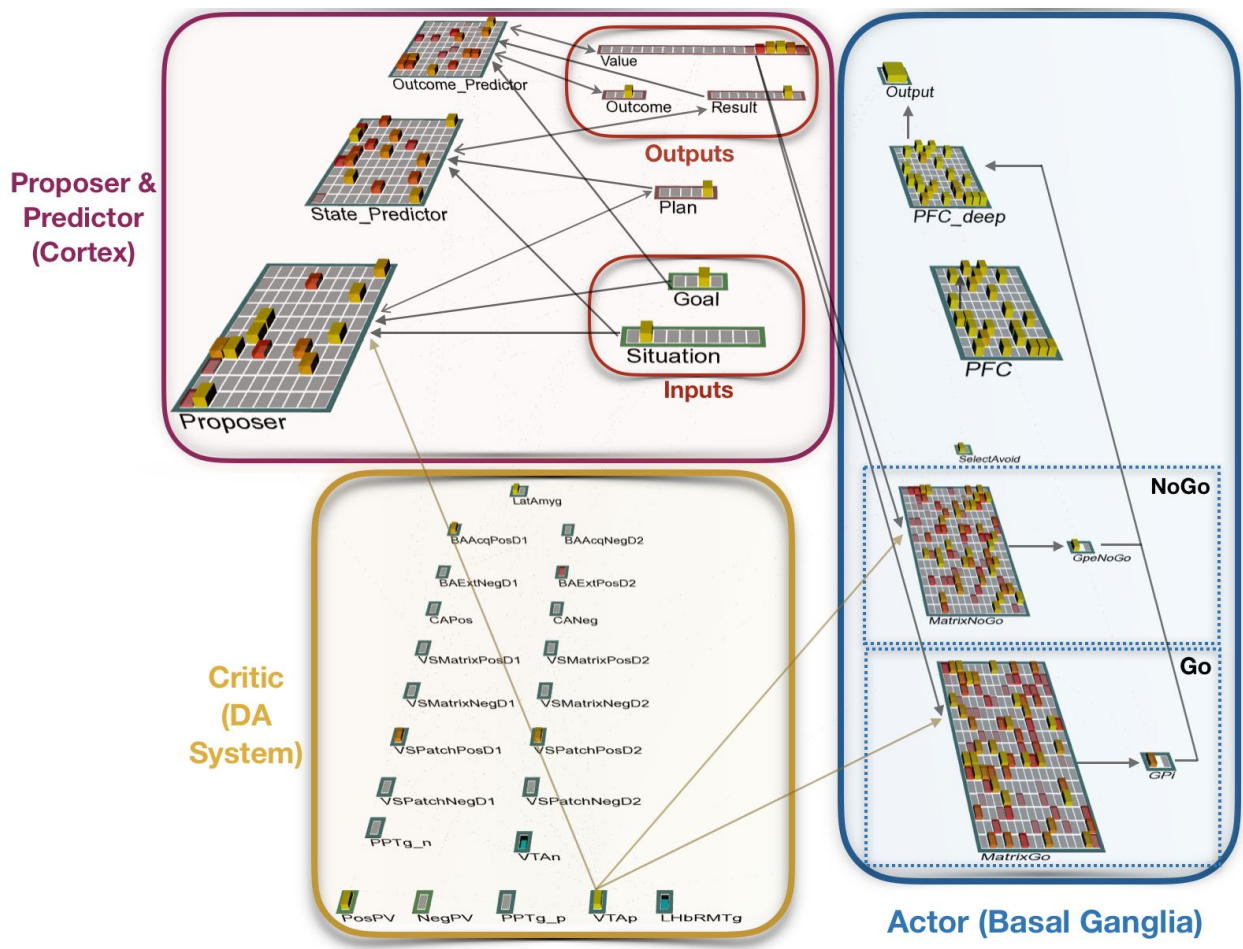

**Figure 4: Model.** The computational model has four broad functional divisions, identified with each element of our Proposer-Predictor-Actor-Critic terminology. The first two are different areas of the cortex (O'Reilly, Wyatte, Herd, Mingus, & Jilk 2013; Munakata & O'Reilly, 2000; O'Reilly et al 2016). The Proposer and Predictor each take the current Situation and Goal, and respectively propose a plan, and predict its outcome given those conditions. Those functions are both distributed across areas of cortex and medial temporal lobe, depending on the specific task domain. The Actor functional division is the basal ganglia loop attached to the relevant areas of cortex (Hazy, Frank, and O'Reilly 2006; Herd et al

2013). It uses reinforcement learning (RL) to "gate" or enhance signals from specific areas of cortex. It thus adds a RL-driven selection mechanism that acts on the associative/predictive representations created by each area of frontal cortex. The final subsystem is the reward prediction system (Hazy, Frank, O'Reilly 2010; Mollick et al, submitted; Hazy, Frank, & O'Reilly, 2007). This system is the Critic; it acts as a teacher of the Actor (basal ganglia) and Proposer systems. It predicts when a reward is likely to occur in the near future, and "discounts" by reducing or eliminating reward signals that it predicts.

**Model computations and biological underpinnings:**

We begin by describing the full model; we tested several variations for this architecture, each of which are described later, in reference to this full model.

The Proposer learns to propose an appropriate plan based on the Situation and Goal inputs. Theoretically, this can be learned from a variety of signals — in the current model we used the dopamine signal, such that connections from inputs to and from the Proposer layer are strengthened or weakened when the Critic produces a positive or negative dopamine signal based on its computed expectations, and the actual outcome. The Predictor consists of two areas and stages of prediction. The State Predictor learns to predict the Result (resulting state) from the current Situation and the Plan currently under consideration. The Outcome Predictor (identified with OFC) layer predicts the Outcome (potential reward, eg., food, social dominance) of that Result (state), and the reward value of that Outcome (it is rewarding if it matches the current Goal). Each of those environmental variables is presented to the network as an input or feedback from the simulated task, each in a localist coding in which a single unit corresponds to a single variable identity.

The Predictor areas of our model pass their predictions to the basal ganglia, which evaluates them in the context of the proposed plan and the current goal. In our main model the basal ganglia layer receives input only from the Value prediction layer of the cortex, which represents the summarized output of this prediction. This single input provides an ideal signal for the basal ganglia to learn from, because after learning the task structure, the cortex is able to make a very accurate prediction of the reward value of the current candidate Plan. Cortical inputs to basal ganglia are known to be diverse and integrative (Haber, 2010), so we think that in reality the basal ganglia often integrates information from multiple sources to make a final decision.

The Actor component in the basal ganglia determines whether to use or reject the current plan. In our current model, the basal ganglia is composed of the Matrix and Globus Pallidus (GP) layers, with the GP layer including the computational roles of the substantia nigra and thalamus in the basal ganglia loops (Hazy, Frank, and O'Reilly, 2006). Go and NoGo pathways (Schroll and Hamker; 2013; Hazy et al, 2007; Collins & Frank, 2014) compete to make a decision, based on weights learned through dopamine reward signals from the Critic component of the model, described below. Thus, the weights for the Go pathway support accepting the current Plan, and those in the NoGo pathway support rejecting the current Plan.

The PFC and Output layers are also elements of the Actor, and represent the downstream effects of selecting the proposed plan. When the GPe layer activates (which in turn happens when the Matrix comparison favors the Go layer over the NoGo layer), the PFC layer is gated into the PFC_deep layer, which in turn activates the Output layer, which we interpret as the model activating the Plan currently under consideration. This architecture is modeled in accord with our existing *PBWM* theory of working memory (Hazy, Frank, and O'Reilly, 2006; Herd,

Hazy, Chatham, Brant, & Friedman, 2014), reflecting the idea that a plan would usually be maintained in an active state so that it can bias further processing accordingly (Miller & Cohen, 2001; Herd, Banich & O'Reilly, 2006). Mechanistically, the PFC contained a second "stripe", not depicted in the diagram, representing the consequences of not selecting the current plan. In this simplified model, we only needed to record the overall Go vs. NoGo gating activity, so all of these additional PFC / Output components are really just place-holders to be developed in subsequent models.

The Critic component of the model is adapted from the PVLV model of reward prediction, detailed in Hazy, Frank, & O'Reilly (2010) and Mollick et al. (submitted). In the current model, we use the *Primary Value* (amygdala and ventral striatum) components, which predict rewards via projections originating in the Output layer, and use this prediction to discount predicted rewards when they occur. Thus, expected rewards generate lower levels of phasic dopamine bursts, and unexpected failures produce phasic dips. These effects are well-documented empirically, as reviewed in Schultz (2013). We did not use the *Learned Value* portion of the model that drives Conditioned Stimulus (CS) dopamine, because that signal is redundant with the Unconditioned Stimulus signal in the current task (since we use a trace learning rule in the Matrix, which allows learning from a dopamine signal to effect the units that triggered the previous action which caused that signal).

The reward-prediction function of the Critic system is crucial because it is the only source of negative learning signals in our primary task — as in many human decision-making domains, there is no explicit punishment; pursuing a plan that does not accomplish the current goal does not cause any direct, physical harm. It is only by predicting a possibility of reward, and then receiving none, by which the agent knows that it has committed an error by missing an opportunity for a valuable reward. Without the reward prediction Critic system in place, the model learns only from success, and as a result, learns to accept every proposed Plan. There is direct evidence that reward prediction signals in the striatum can be influenced by prediction-based computations in this way (e.g., Simon & Daw, 2011; reviewed in Doll, Simon & Dayan, 2012).

**Results:**

**Serial prediction of outcomes**

We tested the model's match to empirical data by performing an analysis similar to one Rich & Wallis (2016) performed on their neural recording data from monkey OFC. In our analysis, we looked at activations in the Value layer on a cycle by cycle basis, correlating its activations with the idealized activation for a good and a bad option. This shows how the value prediction of the currently considered plan evolves over time, including how it shifts as the plan the model is considering shifts after rejection of a plan. Both seem strongly indicative of serial performance. While this is not surprising, as we designed the model to perform a serial analysis of options, the match between this data and of Rich & Wallis lends support to their hypothesis that their monkeys were serially considering response options, and that their OFC (and likely, other linked, contributing brain systems) was predicting outcomes of one possible response at a time.

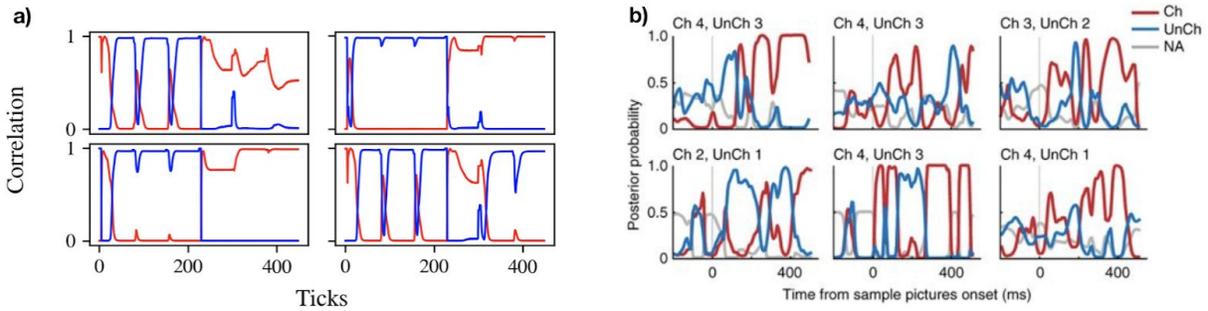

Figure 5: Model Value representations over time, illustrating the serial consideration of one action at a time, in model and empirical data. a) Analysis is a normalized cosine similarity to canonical Value representations. In this trial, the model considered two Plans whose outcomes did not match the current Goal, and the Predictor component correctly predicted a low Value, before correctly predicting the rewarding Outcome of the third Plan it considered. b) Similar analysis of monkey OFC representations during decision-making. Normalized cosine similarity to average of neural activity at time of decision (Rich & Wallis 2016). Each analysis shows a relatively clean representation of the predicted outcome of one option at each moment in time, indicating a serial prediction computation.

**Basic Model Results**

We report several results from our model, demonstrating its basic functionality and accounting for central aspects of the empirical data. We first discuss performance of the primary ("full") model, discussed in detail above, then compare that model to several variants. Each comparison illustrates a different computational aspect of the full model's performance.

Each of our results were taken from 50 runs using different random starting connection weights, train/test splits, and ordering of trials and random selection of Plan for consideration for the no-Proposer models.

The State Predictor learns to correctly predict a *Result* (93.1% +- 0.4) given an input of a *Situation* and a *Plan*, while the Outcome Predictor learns to predict both an *Outcome* (87.9% +- 0.9) and a *Reward* (98.0% +- 0.05) given a predicted *Result* (from the State Predictor layer) and a *Goal*. Because their roles are each split out separately, this becomes a relatively easy learning task. Note that the learning task is made harder by the fact that these layers only learn when an option is selected, so that sampling is uneven; as the model learns to select correct options, it ceases selecting and learning about non-rewarding options. This bias toward exploitation versus exploration can negatively affect learning. It can prevent learning about new combinations of Situation and Plan, and causes the cortical Predictor layers to overwrite their correct predictions about non-rewarding (and so decreasingly-sampled) Plan-Situation combinations until they are incorrectly predicted. As a result, the basal ganglia Actor starts to select these non-rewarding options, which then triggers self-correcting new learning. Thus, the model has an intrinsic tendency to titrate between exploration and exploitation, and by approximately maximizing reward in the short term, it never achieves maximal performance. For example, the full model averaged between epoch 500 to 600 over 50 batches chooses an optimal plan in 93.0 +- 0.5% of trials on the training set.

**Computational shift and speedup with experience**

Our model shows some transition from slower, more prediction-based to faster, more prediction-free computations. This transition has long been a topic of interest in psychology under the terms *controlled versus automatic behavior* (Shiffrin & Schneider 1977; Cohen, McNaughton, & McClelland, 1990). It has also been addressed in terms of a shift between *goal-directed* to *habitual* behavior (Tolman, 1948; Tricomi, Balleine & O'Doherty, 2009); however, that distinction does not map directly to prediction-based vs. prediction-free computations, as illustrated by the fact that our prediction-free Proposer component takes the current Goal into account and learns to produce candidate Plans that accomplish that goal. Because it selects a candidate Plan based on its weights, learned from the relationship between Situation, Goal, and previously performed Plans, the Proposer model component is performing a prediction-free computation.

As the Proposer component learns, we found that it reduces the number of Plans that the full model considers, and so saves substantial time, as shown in Figure 6b and 6c. Thus, to the extent that the Proposer can propose a good plan early in the consideration of options, it can significantly reduce the overall decision-making time. We think this speedup is made more dramatic in some cases by not waiting for the Predictor's contribution at all once the Proposer is sufficiently successful in proposing good Plans; we model that possibility in the no-Predictor variant model architecture (while explicitly modeling the mechanisms that control that shift await future work).

Our full model generalized well but not perfectly to the held-out test set (62.9 +- 3.3%). This generalization performance seems surprisingly poor given the model's computations. Because it separately predicts Outcomes given Plan and Situation, and Reward given Outcome and Goal, it should be able to produce near-perfect generalization to the test set of withheld Situation-Goal combinations. In contrast, the no-Proposer model, in which plans are considered by random draw instead of learned selection, performed very well on the generalization test (93.7+-1.2%) (Fig 6a). We found that the full model sometimes generalizes incorrectly despite thorough training because the basal ganglia has learned to accept whatever plan the Proposer proposes, as the Proposer often proposes the correct plan on the first try (60.5% +- 1.5% after 500 epochs of training). Because the Proposer is never trained on the held-out test set of Situation-Goal combinations, it proposes the correct plan at chance levels in the generalization test (30.4% +- 3.2), and yet the basal ganglia will have an increased tendency to approve the first candidate plan. This produces worse performance on generalization vs. training (62.9% +- 3.3 on the held-out test set, vs. 93.0% +/- 0.5 on the training set; 6a). This behavior of the model suggests interesting empirical tests of biased training sets that could reproduce similar effects as found in this model.

Figure 6

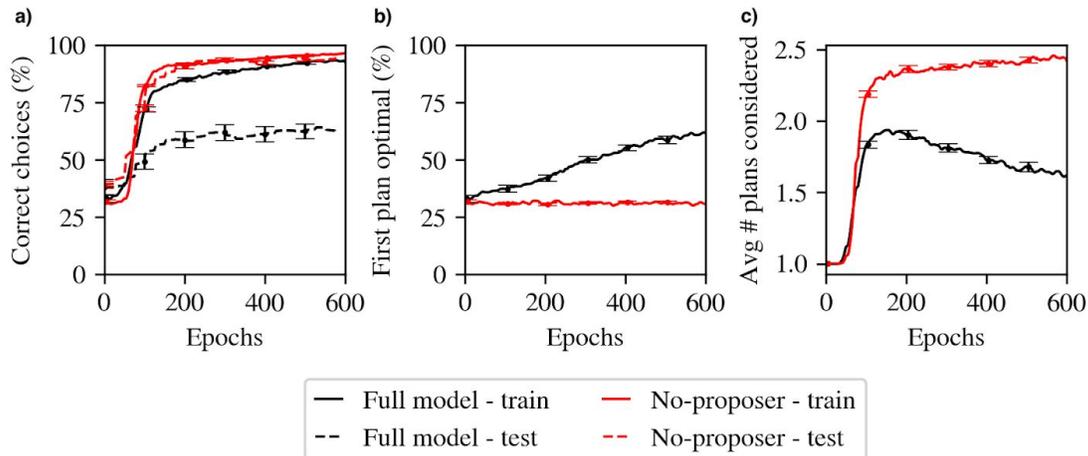

**Figure 6; Full model speedup and no-Proposer model a)**, test behavior for full (black lines) vs.no-Proposer (red lines) models. The no-Proposer model considers a randomly-selected candidate plan on each time step. It performs even better than the full model, and performs almost as well on the generalization set as the training set. **b)** Fraction of first candidate plan correct for the Proposer layer (black line) vs no-Proposer (red black line). The Proposer learns across the course of 600 epochs to select rewarding candidate plans. **c)** Number of candidate plans considered. The full model considers fewer plans per decision, on average, since its Proposer component learns to select useful plans to consider first. Thus, the Proposer portion of the model allows it to perform faster as it learns, at the cost of worse generalization performance.

**No-Predictor Model**

Our primary, full model addresses how the brain might perform prediction-based decision-making. To address how the human brain might perform faster, but less accurate decision-making, we ran two comparison models, one with no explicit prediction, and another which predicts only the Value of a Plan-Situation-Goal combination, without predicting the Result or the Outcome.

In the first comparison model, we simulated the original hypothesis of Daw, Niv, & Dayan (2005) that the basal ganglia performs prediction-free decision-making. This subcortical model-free version of the model uses only the basal ganglia to accept or reject each plan, with no prediction from the Predictor component. Without any contribution from the Predictor component, the Value layer has no activation. Thus, it was necessary to change the connectivity of the model by introducing a direct connection from the *Plan, Situation and Goal* to the *MatrixGo* and *MatrixNoGo* layers, and removed the connection from the *Value* layer. This model performed poorly with default parameters (32.5% +- 0.5% on the training set, around the empirically determined chance performance level of 32.1%+-0.6%). This appears to be the result of a bias toward accepting plans when the basal ganglia matrix layer has more input layers. We adjusted the threshold level in weighing the matrix Go vs. NoGo activities, from .1 to .5, and saw better performance of 50.47% +- 0.7% for training. This variant performed at 29.1% +- 1.9% on the testing set, as expected, since it does not segment the task into predicting Results and Outcomes, and so should not be able to perform above chance on the holdout test set. When we added a punishment value of .5 to all Outcomes that did not match the current goal, the model's performance on the training set improved to 70.6% +- 1.5%; this appears to be the result of further reducing the Go bias early in learning. This variant still performed at or below chance levels on the testing set, 27.0% +- 1.5%. Our model thus predicts an upper bound on the types of decisions that can be learned by the basal ganglia without predictive input from the cortex, in line with other predictions from other theories of model-free decision-making.

We agree with the hypothesis that some decisions are made in a model-free mode that relies primarily on basal ganglia (Daw, Niv, & Dayan, 2005). However, we would predict that even in those situations, the basal ganglia are making their decision using highly processed cortical representations of sensory/state information as input, so cortex participates heavily even in "prediction-free" decisions. It may be difficult to fully eliminate any form of cortical model-learning from even the most basic forms of human decision-making, consistent with evidence reviewed by Doll et al (2012).

One advantage of prediction-free decision-making is that it should allow faster performance, because the basal ganglia need not wait for a prediction (or whole series of predictions) from cortex or medial temporal lobe. Consistent with this model, Oh-Descher et al (2017) have observed a shift from cortical to subcortical activity when time pressure was increased, accompanied by a shift in decision style to use a simpler set of criteria.

**Value-only Predictor model**

People may also switch to strategies with less outcome prediction when the task structure becomes more complex, increasing the difficulty and demands of explicitly predicting outcomes (Kool, Cushman & Gershman, 2018). We illustrate another such possible strategy with our next model variant. In this hypothesized decision-making model, the cortex participates in model-free, but prediction-based, decision-making. In this hypothesized strategy or model, the computational power of the cortex is used to learn and predict which actions will be successful, but without making specific predictions about outcomes. This model variant instead makes a simpler, one-step prediction of the value of a Plan in a given Situation. This type of prediction is model-free according to the commonly used definition, since it does not include a prediction of a specific outcome. The prediction should be faster (because it demands fewer cognitive steps) but less useful in novel situations.

This version of the model therefore learns about the value of a plan in accomplishing a given Goal in a given Situation, but without using a serial process to predict specific outcomes, in a specific order, as our primary model does. Instead, the two cortical layers work in series, to produce a more powerful "deep" network with two hidden layers, which takes in all of the relevant information, and produces only a predicted Value as an output. We think that humans may sometimes use the cortex and basal ganglia in this technically model-free, but prediction-based way. This computational approach brings the computational power of cortex to bear, without requiring individual, time-consuming steps for each specific predictive step necessary to arrive at a likely outcome in a complex task.

This model performs well on the initial training set (88.4 +- 0.8% optimal Plans chosen after training), but it does not generalize much, if any above chance level on the testing set (27.7+-4% vs. an empirical chance level of 32%+-0.6 (Figure 7) (Measured as average performance during an all select pre-training with random plan choice). This illustrates one key advantage of prediction-based decision-making, when predictions are organized into separate steps: it can produce the correct decision on the very first encounter with a new combination of known elements (e.g., starting a known maze with the reward in a new but known location, etc), because each element's outcome is predicted separately.

This version of the model is at a second disadvantage: without making specific predictions, it cannot be organized to learn separately about the two steps of prediction for this task. Thus, to give this approach every chance to succeed at generalization to the held-out task combinations, we increased the size of both cortical value-prediction layers from 100 to 400 units, and gave this version of the model 50 epochs of pre-training on predicting outcome values. During this pre-training, the model learned about the values of uniformly sampled random Plans for each Goal and Situation combination, before the Proposer, Actor, and Critic components were allowed to choose and learn.This biases the experiences of the Predictor component.

Without this pre-training advantage, the model performed much worse on the primary task (76.9 +-1.3%); because of this poor general performance in the absence of pre-training, we do not draw conclusions from its even worse (roughly at-chance) generalization performance.

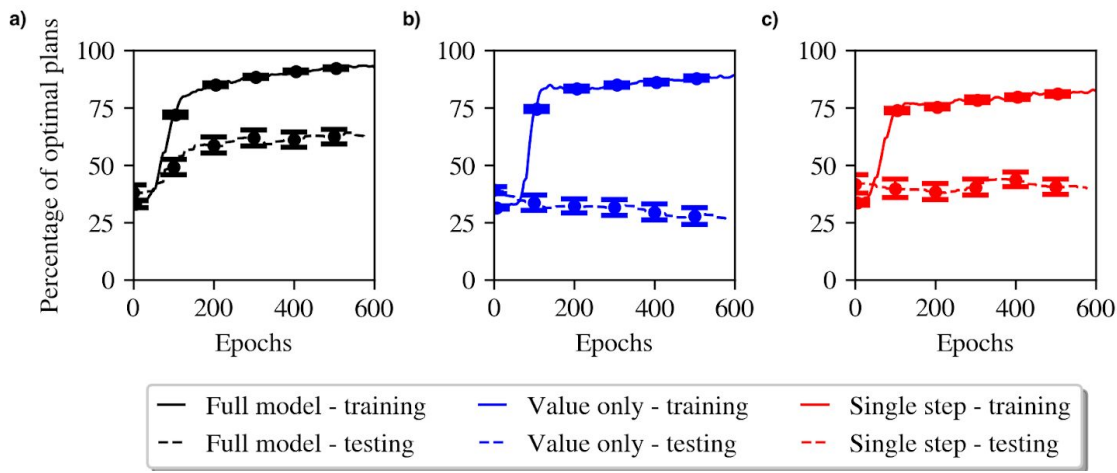

**Figure 7: Reduced Prediction Model Results.** The full model performs at around 95% accuracy on the full task (that is, rejecting poor Plans until choosing a Plan that achieves the current Goal for each Situation/Goal combination). It performs reasonably well on the testing set of reserved, never-before-seen combinations of Goal and Situation, tested every 25 epochs). Two versions of an unstructured, single-step prediction model performed much worse on the testing set. Each of these models used the two predictor layers in series, without the intermediate State Prediction step trained independently. The Value-only model was trained to produce only the value, while the single-step model produced all of the same predictions as the full model, but without separating those by layer - one layer acts as a hidden layer, and the output layer predicts State, Outcome, and Value. This very different performance on the hold-out test set illustrates one key advantage of prediction-based processing with structured predictions in discrete steps: it can produce correct decisions on the very first experience with new combinations of known causal factors.

**Unstructured single-step Predictor comparison model**

Humans appear to be capable of decomposing their predictions of outcomes into multiple discrete steps. For instance, in planning our day we may predict that taking the northbound freeway will get us downtown, then, in a separate cognitive step, predict that being downtown will allow us to meet a potentially valuable contact for lunch near where they work. This decomposition of problem space into sensible subcomponents offers substantial computational advantages (at the likely cost of slower performance). Our primary model breaks its prediction into two steps, *State* and *Outcome*. To illustrate this computational advantage, we ran a comparison model that does not break the prediction into discrete steps.

In reality, we think that humans can use a flexible and unlimited number of predictive steps, but for simplicity our computational model always uses two steps. The State Predictor layer predicts a Result, and the Outcome Predictor layer uses that prediction to predict the Outcome linked to that Result (and the reward value that results from that Outcome in combination with the current Goal). To illustrate this advantage, we ran another comparison model that does not explicitly separate those two predictive steps. In this version, there are still two cortical layers, but they are organized in a strictly serial manner: the first layer receives all inputs (Plan, Situation, and Goal) and projects to the second layer, which projects to each prediction layer: Result, Outcome, and Value. Thus, this comparison model is a standard deep network with two hidden layers. The cortical layers are larger (400 units vs. 100 in the main model), to give this model a better chance of performing well. When the model is so arranged, it somewhat worse than the full model on the training set (81.7 +- 1.0%), and dramatically worse on the generalization tests (40.4 +- 3.6%) (Figure 7). (When the two hidden layers are held to the same size, the model performs slightly worse - training: 74.2 +- 1.4; testing: 29.3 +- 3.6.)

This result is interesting, because it is certainly possible for a neural network to decompose a complex problem into its components and so achieve good generalization through error driven learning. This computational principle has been demonstrated by the recent success of deep networks on visual object recognition (Krizhevsky, Sutskever, & Hinton, 2012; Wang et al, 2017). Such learning may even produce internal predictions of outcomes in the hidden layers, despite being trained only to predict their value, as we address in the Discussion section. However, those networks receive millions of training trials, while biological brains experience only handfuls to thousands of decisions in each domain (Lake et al, 2017). Creating a-priori divisions in predictive steps can reduce the amount of training needed for generalization to new situations dramatically, if those divisions match the structure of the real world. Exactly how those predictive steps are created to match the world's structure is an outstanding question, but it is likely to include directing attention to different types of outcomes.

**Applications to Risky Decision-Making:**

One notable application of a mechanistically detailed theory of human decision-making is in understanding how humans make bad decisions. Taking large risks, such as driving while intoxicated, risky sex, and dangerous drug use are all areas in which bad decisions produce enormous societal and personal costs. Risky decisions in political and military domains can produce even worse impacts. While the model as it stands cannot fully describe the complex processes by which humans make such decisions, it does still offer some potential explanations of factors leading to more risk-averse or risk-tolerant styles of decision-making.

Our comparison between the full model and the no-Predictor variant shows how the human brain can support two distinct approaches to making decisions; a fast approach, without explicit predictions of outcomes, and a slower but more accurate prediction-based approach. This speed/accuracy tradeoff between model/prediction-free and model/prediction-based computations appears to be a common supposition. If this is correct, one major factor in mitigating risky decision-making is to use interventions that encourage a careful, prediction-based approach when decisions may have serious consequences. While this suggestion may seem obvious, it is not clear that existing interventions have explored this strategy.

We performed additional modeling of another potential factor in risky decision-making: individual differences in propensity to approach and avoid. These individual differences have been characterized as the two opponent systems, the Behavioral Inhibition System and Behavior Approach System (BIS/BAS). There are many individual genetic and environmental influences that could affect approach and avoidance behaviors; we manipulated a fairly basic parameter controlling our model's Go vs. NoGo behavior.

To address risky decision-making, we modified our task to more closely reflect real-world situations in which risky behavior has been identified and studied. In most such real-world domains, there are rare, highly negative outcomes, balanced against more frequent, smaller rewards. To approximate that profile, we modified our task and rewards to produce more small rewards and a few large punishments. We used the same basic task structure, but made three goals (instead of one) rewarding for each trial, and added stochastic large punishment when the model selected the remaining single bad outcome for its current goal on any trial. We rewarded the model (with a .2 reward value) for achieving an Outcome that matched any of those three randomly selected Goals. We stochastically punished the model (with a punishment of 1) on 25% of the failing trials (in which it arrived at an Outcome that matches the one currently invalid Goal). This produced a total base rate of 6.25% (1 out of 16) punishment trials and a base rate of punishment to reward amounts of 12.5% punishment.

We then manipulated the model (our full model, with all components as depicted in Figure 2,). Differences in reward learning and decision-making have been observed in many studies when drugs and optogenetics have been used to differentially strengthen the Go (D1-receptor-dependent) and NoGo (D2-receptor-dependent) pathways in dorsal striatum (corresponding to the Actor basal ganglia model component of our model). (Frank, Seeberger, & O'Reilly 2004; Moustafa, Cohen, Sherman, & Frank 2008; These results show the expected difference.  in line with published experimental results suggesting that strengthening D1 pathways leads to increased responding to rewarding options, while strengthening D2 pathways leads to increased avoidance of bad options (Frank, Moustafa, Haughey, Curran, & Hutchison, 2007). This is also consistent with results showing that driving D1 neurons in the Go pathway directly causes motor actions (Sippy, Lapray, Crochet, & Petersen, 2015).

We simulated this by manipulating a gain factor moderating the competition between those pathways in our simulated Globus Pallidus internal segment (GPi layer), the final step in the basal ganglia loop that decides whether to use the current candidate plan (Go) or reject this plan and consider a different candidate (NoGo). Strengthening the D2-driven NoGo pathway in our model relative to the D1-driven Go pathway leads to a more cautious, less risky behavioral profile in which fewer options are selected overall (Figure 8, below).

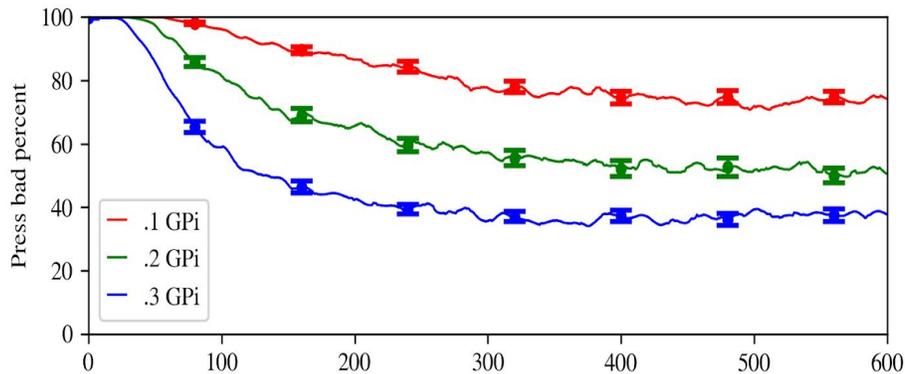

**Figure 8: Go vs NoGo pathway Manipulations.** Risky decision making model with increasing strength of NoGo pathway (modeling D2 receptors in dorsal striatum) Models with higher relative D2/NoGo pathway strength accept fewer bad options, creating an overall less risky behavioral profile.

**Vicarious Learning for Risky Decisions**

In many domains, including risky decisions, people seem to base their decisions not on their own experience, but on what they have learned about the experiences of others. They might simulate the experiences of others in enough detail to produce a relatively complete physiological response to good or bad outcomes, including dopamine release; in this case, our model of learning for decision-making would function identically whether experiences were personal or vicarious. Alternately, in some cases people may learn information in the abstract, without experiencing the physiological responses.

We simulated such a learning simulation by separately training the Actor portion of the model (basal ganglia) to accept or reject Plans based on inputs from the Value layer, using a training set of good and good and bad outcomes chosen with equal probability. This phase of training modeled early learning of good and bad experiences in a variety of domains, and training the Predictor component (cortex) on our risky decision task, described above, but in this case with a reward value of .5 for each correct choice (the choice of .2 reward for the Go vs. NoGo manipulation was chosen to show behavioral differences more clearly, but using .5 reward also showed smaller differences). This training simulated vicarious experience with the abstract semantics of this task, for instance, hearing about others' successes and failures in gambling, inebriated driving, risky sex, etc. We then tested the model on making decisions on that task, but with all learning turned off.

This test provided a measure of model behavior on risky decision in a domain that has only been experienced vicariously (e.g., the first time a teenager chooses whether to ride with an inebriated driver). We demonstrated that a model trained on mostly negative outcomes made many fewer risky decisions than a model trained mostly on positive outcomes (8.1% +- 1.3% chance of choosing a bad plan for the negatively over sampled model vs 50.3% +- 3.3% for the positively over sampled model). This result follows straightforwardly from the model's distinction between Actor and Predictor components; this demonstration illustrates how those components map to abstract knowledge, and full experience (whether direct or simulated) of rewarding and punishing outcomes.

**Discussion:**

We have presented a specific computational implementation of our *Proposer-Predictor-Actor-Critic* model for how the core loops between the frontal cortex and basal ganglia can support even complex human-level decision-making unfolding over multiple sequential steps. At each such step, the Proposer leverages constraint-satisfaction and learned domain knowledge to activate a plan, which is sensitive to the current external and internal state and situation. This proposed Plan can (but does not always) drive the Predictor to generate likely outcomes at multiple levels of analysis, with varying levels of time and effort that could be invested to develop varying levels of detail and accuracy in these predictions. This information feeds into the basal ganglia Actor system, which makes an overall Go vs. NoGo evaluation based on a history of dopaminergic learning, which in turn is driven by the reward-prediction-error signals from the Critic. This model builds on extensive existing work on action selection mechanisms in the basal ganglia, and the overall Actor-Critic computational framework of Sutton & Barto (1981). In addition, it provides an alternative to the strict model-based/model free dichotomy of Daw et al (2005); our model performs both types of computation, but does not separate them into two distinct systems.

We showed that our model can learn to accurately propose and predict outcomes in a multi-dimensional decision-making task that captures important aspects of real-world tasks, and that using predictions of specific outcomes produces better performance. Furthermore, the model was able to generalize its knowledge to novel task configurations that it was never trained on; this generalization ability also depends critically on its predictive-learning components. As training proceeded, the proposer aspect of the model became much better at generating appropriate plans, and this affected the balance of exploration vs. exploitation, while also significantly speeding up decision-making performance with expertise. Both of these phenomena emerged out of the more basic properties of the model, and provide important first-principles predictions about the dynamics of learning in people and other animals.

Finally, our model accords with the data of Hunt et al (2018), Rich & Wallis (2016) and others who have found evidence that the OFC and related brain systems encode reward predictions about whatever option is currently attended, relative to other options. Hunt et al (2018) also interpret their recordings from ACC to indicate that the ACC is also considering a single option, the best one encountered thus far, and accepting or rejecting that option. Because the monkeys gave a left-vs-right manual movement response, we find it somewhat surprising that the ACC did not act as a parallel accumulator of the estimated relative value of each action; the ACC has a good deal of real estate devoted to manual left-vs-right movement, and their task was quite well-practiced.

Their data is not conclusive, as they did not perform strong tests for neural representations of the other options, but their finding of a large accept-reject signal in even a relatively simple task may indicate that, even for relatively simple and fast decisions, people do not accumulate relative values for multiple options in parallel, but make a fully serial consideration of one option at a time, despite the non-optimality of doing so in purely computational terms. This would be consistent with some theories of decision-making (some reviewed by Hayden, 2018); inconsistent with others (some reviewed by Turner, Schley, Muller, & Tsetsos, 2018); and entirely orthogonal to many aimed at the psychological level, with less specification of mechanisms and specific computations.

In the remaining sections, we consider the relationship between our framework and other existing frameworks, and associated empirical data, and then enumerate a set of testable empirical predictions from our model.

**Relation to theories of model-based vs. model-free decision-making**

Overall, our framework has a significant amount of overlap with the model-free vs. model-based framework (MFMB) originally elaborated by Daw et al (2005), and widely discussed in the subsequent literature. Both frameworks include an essential role for predicting future sensory and other states based on internal "models" of the environment, along with a critical role for dopamine-mediated learning to select "good" vs. "bad" actions. However, whereas the MFMB framework is based on a dichotomy between these two types of processes, our framework emphasizes their synergistic interactions within the context of the characteristic fronto-striatal circuit replicated across many frontal areas. The cortical and basal ganglia components of this circuit each contribute unique and important computational functions to the overall decision-making process, supported by their unique neural properties.

Thus, we think that the dichotomy envisioned in the MFMB framework is typically much more of a continuum, with model-like elements likely to be involved in many cases to varying extents. Our model, and this proposed continuum, is consistent with key findings from fMRI studies by Daw, Gershman, Seymour & Dayan (2011) and Simon and Daw (2011), where the striatum seemed to be involved in prediction-based decision-making, in that its activations correlate with predictable outcomes. In addition, Kool, Cushman, & Gershman (2016) found that people typically exhibit a complex mixture of both model-free and model-based profiles.

**Predictions of the model**

One important distinction between the current theory and the majority of work on human decision-making is our reliance on detailed depth-recording and tractography data. The component models were created based largely on animal work. While those animals are not trained to perform tasks as complex as those we address and model, that work provides vastly more detailed information on neural computations than can be obtained from humans. Our central argument here is that the same circuits that perform action-selection in animals are sufficient to explain complex decision-making in humans, when they are employed serially, and have access to the more abstract representations available in the human brain. Our model makes strong predictions about the interaction between cortex and basal ganglia, and neuroimaging and neuropsychological approaches with human subjects should be able to test those predictions and so falsify or support this theory.

Our primary predictions are the computational divisions-of-labor we have emphasized throughout: cortex produces a candidate plan, other cortical regions may produce a model of the predicted outcomes, and connected loops of basal ganglia make a final decision to use that candidate or reject it (at least for the moment) and consider another candidate in a serial, iterative process. Perhaps our most central prediction is that human decision-making, and the predictions upon which it relies, work serially in relatively novel domains. It is somewhat challenging to test this prediction. Tracking precise timing of representational content in humans is challenging with current neuroimaging techniques, and allowing subjects to select their own ordering of sequenced steps makes this problem more difficult. Interpreting detailed animal recording data can be difficult; for instance, the neural recording data of Lorteije et al 2015 were re-interpreted to fit models of one-step parallel process (Hyafil & Moreno-Bote, 2017). However, animal data (e.g., Hunt et al 2018; Rich & Wallis, 2016), and clever human behavioral designs (e.g., using priming at different times during decision-making in concert with self-reported ordering of predictions) can bring evidence to bear on this prediction.

While the current model has two prediction steps hard-wired, we think that another similar decision may be required to perform each predictive, model-creation step. These sub-decisions may be performed by the same Actor component of basal ganglia, which might use the same dopamine reinforcement signal to learn to delay a final decision long enough for a cortical prediction to play a role, or to accept the Proposer's first Plan to perform quickly under time pressure. It is also possible that the decision to perform more predictive steps may use an anatomically distinct but functionally analogous loop of the same canonical circuit, a loop of basal ganglia associated with an area of cortex that uses domain-appropriate learning to predict specific outcomes. This anatomical and computational question awaits further computational and empirical work. These possibilities are certainly differentiable empirically, but testing them with existing methods would be challenging.

While existing theories and evidence suggest that OFC does often predict reward value and stimuli that are closely linked to reward (reviewed in Rudebeck & Murray, 2014), we assume that predictions of Results will occur in different brain regions for tasks in different domains, even when their task structure is identical. While existing evidence is consistent with this prediction, it is currently insufficient to eliminate the alternative hypothesis that predictions of outcomes are always made in the same brain regions, regardless of domain. For instance, state prediction error signals have been observed in the intraparietal sulcus and in several areas of lateral PFC when people either observed or performed a spatially-arranged state selection task (Gläscher, Daw, Dayan, & O'Doherty, 2010). This prediction can be further tested with relatively straightforward neuroimaging methods.

**Relation to other theories of decision-making**

There are a number of other existing theories of how cortex and basal ganglia contribute to decision-making. Our model and theory draw from those previous theories, but has distinctions from each.

Dayan (2007) has explored some consequences of a theory much like ours for sequencing complex behavior. He implemented a simple model of reinforcement learning, which he identified with BG, PFC, and hippocampus, and trained it to produce multi-step behavior. He identified the usefulness of such a system in producing complex behavior based on verbal instructions. That work was based on the neural and biological model of O'Reilly & Frank (2006), upon which our current theory is also based. Although it does not directly address decision-making, that theory is focused on how the basal ganglia's gating of information into working memory can produce arbitrarily complex behavior. Thus, this theory is very closely related to the current one.

Solway and Botvinick (2012) present a computational theory of prediction-based and prediction-free decision-making that is closely related to ours. Their model similarly includes PFC, BG, OFC, and amygdala, performing predictions for both outcomes and reward value, and includes an instantiation as a learning neural network model. The biggest difference between our models is that theirs performs action-selection as a parallel process, whereas ours predicts outcomes for only one option at a time. This is a central feature of our model; we believe that parallel plan selection can produce fast and useful actions when there is sufficient experience with a specific decision, but that a serial prediction process is a key component of human generalization of knowledge in complex domains. Most theories that take detailed empirical data (e.g., animal single-cell recordings) into account similarly propose a parallel consideration of multiple options. Our theory holds that adapting this system to consider options serially allows humans to make decisions in more complex and novel domains.

Another interesting difference is that their theory focuses on the Bayesian inversion of a model of outcomes and task-space; this roughly equates to backward-chaining from desired outcomes to actions, while our theory currently only addresses forward chaining. It seems likely that humans can perform both types of chaining; the circumstances favoring and computational constraints surrounding each strategy remain a question for future work. Solway & Botvinick's model also differs from ours in identifying the hippocampus and medial temporal lobe (MTL) with the outcome-prediction component. We think the MTL is probably involved in both proposing plans and predicting outcomes when there is little experience with the task domain, so that one-shot learning is necessary. The current task and model include cortex in both Proposer and Predictor roles, for simplicity.

Collins and Frank (2014) present a model with substantial overlap with our model of the basal ganglia (and indeed they share a common ancestry; Frank, Loughry, & O'Reilly, 2001). Their OpAL model accounts for incentive effects of dopamine that our model does not, and proposes a theory, in accord with empirical evidence, of how background dopamine levels can focus basal ganglia decision-making on opportunities vs. rewards in a flexible and useful way. Like Solway & Botvinick (2012), and many other theories based in part on neural networks and animal data, their theory addresses a parallel action selection process, which we think is employed for well-practiced decisions (like many laboratory tasks), while ours proposes a serial process for more novel task spaces. As noted above, the current theory proposes that the same mechanisms that perform well-practiced action selection in parallel in animals have been adapted to work serially for human decision-making in complex domains.

Daw & Dayan (2014) give another computational theory of prediction-based and prediction-free decision-making. They focus on the computational merits of prediction-based vs. prediction-free decisions. They propose that prediction-based decision-making relies on sparse sampling of problem-spaces, which is critical in problem spaces which are too complex to allow for a full start-to-finish prediction of every possible action. This is a critical computational requirement, and our model partially but incompletely addresses this point. Our model does not directly confront this issue, since it currently works only in a limited problem-space in which a full prediction can be made for each candidate Plan. However, our model does include an element that can help in addressing this problem. Our Proposer component selects plans for more detailed outcome prediction, in a fast, parallel neural constraint satisfaction process. This efficiency also contribute to effective sparse sampling of a plan space to focus on areas that are likely to be productive.

Daw & Dayan (2014) propose a different mechanism contributing to this same computational efficiency: prediction-free estimates of state value are employed in complex tasks to avoid the time commitment of following every model through task-space to a conclusion. We propose that the basal ganglia makes prediction-free decision at every step of model building. Our current model does not perform this function, but we intend to capture this in future extensions of the model. Similarly, we propose that the decision to use cortical, prediction-based reasoning is itself made by the same canonical circuit of linked cortical and basal ganglia loops. Consistent with this prediction is evidence that deciding on task strategy activates the frontopolar cortex and inferior lateral PFC (Lee, Shimojo, & O'Doherty, 2014).

Daw, Niv, & Dayan (2005), followed by many others, propose a categorical distinction in which PFC performs prediction-based computations and basal ganglia performs prediction-free decisions. In our model and theory, basal ganglia contributes to both types of decisions, and cortex may also (see the cortical prediction-free model section). Subsequent empirical work has called this strict separation of systems into question. Cortex now appears to be heavily involved in habitual behaviors (Ashby, Turner, & Horvitz, 2010), and there is strong evidence that basal ganglia plays a critical role in higher-level cognitive function (e.g., Pasupathy & Miller, 2005; Balleine, Delgado, & Hikosaka, 2007), and in goal-directed behavior (Yin, Ostlund, Knowlton, & Balleine, 2005). Our model builds upon that evidence. Our cortical Predictor component performs an optional extra step, adding information to the prediction-free system, in distinction from that proposal of two parallel and competing systems. This theory, and follow up work (e.g., Daw & Dayan 2014) (along with most theories of decision-making) do not directly address the parallel vs. serial distinction upon which we focus, although their theoretical treatment of human decision-making seems consistent with assuming a largely serial process, in which each prediction adds a nontrivial time cost.

Buschman & Miller (2014) present a theory with substantial overlap but substantial differences from ours and that of Daw & Dayan (2014). They focus on a related but separate distinction: BG learns concrete associations quickly, while PFC slowly learns more abstract concepts for decision-making. Our model does not currently address this distinction; doing so is a promising avenue for future work. Buschman & Miller propose several possible computational advantages of the interaction between PFC and BG, none of which map closely to our Proposer-Predictor-Actor distinction. In particular, they propose that such loops may allow for stereotyped sequences of actions or thoughts to be strung together, in analogy to the well-studied role of BG in contributing to sequences of motor actions. Our model is consistent with this role, but does not currently address it. They also make the suggestion that PFC's ability to capture abstract concepts allows the BG's action-selection to work in more abstract domains. While we propose a collaborative model of action selection between PFC and BG, we agree that this expansion of animal action-selection to complex human decision-making relies on the learning and use of abstract representations in PFC, another computational advantage which we do not directly explore in the current model.

Our proposed mapping from prediction-based vs. prediction-free decision-making to anatomy is compatible with that proposed by Khamassi & Humphries (2012). They propose that dorsomedial striatum participates in prediction-based action selection, while dorsolateral striatum participates in prediction-free action selection in rodents. They do not specify a serial or parallel mechanism, and do not specify how those systems interact, while we focus on those specifications.

Koechlin and Hyafil (2007) also present a theory broadly consistent with ours, but specific to cortical contributions; they do not address the role of basal ganglia or dopamine. They focus on human anterior prefrontal cortex (APFC) and review evidence showing its importance in complex, branching decision-making tasks. Donoso, Collins, and Koechlin (2014) present a related theory of how APFC is involved in strategy testing in complex tasks. These are both consistent with our theory, although our current model does not directly address the distinction; we propose that the same circuits we outline here are also at work in the APFC, and the same serial process is used to select strategies (or plans in our terminology) based on their previous success.

Our model and theory are also broadly compatible with Botvinick and Weinstein's (2014) review of work in hierarchical reinforcement learning. We make similar arguments for the computational advantages of making decisions hierarchically: selecting a broad plan, then subgoals, and only finally selecting actions. While our theory is compatible with that work, our focus here is on outlining the neural mechanisms that instantiate that computational process, and our current model does not progress through such a hierarchical decision process. Expanding the model to perform such a process is another topic for future work.

A great deal of work has been devoted to mathematical theories of decision-making. Because those theories make no contact with detailed neuroscience data, there is (so far) little contact between the current theory and that mathematical level of analysis. Because it allows for a variety of decision-making strategies, composed of different cognitive steps, it could be used to match results from many of the wide variety of models that are still being debated (Hastie & Dawes, 2010). We view those theories as working on a separate but complementary level to this one.

One notable exception is the drift-diffusion framework, as it has been closely related to anatomy and neural data. The mathematics of drift diffusion have been argued to closely characterize neural firing in relevant areas of cortex during perceptual decisions (e.g., lateral intraparietal lobe during a task based on movement; Hanks, Ditterich, & Shadlen, 2006). Neural activity in basal ganglia have also been shown to closely map to a drift diffusion model (Ding & Gold, 2013), and it has been argued that the basal ganglia's structure matches an extended version of drift diffusion that takes into account the evidence accumulated for other options to achieve optimality under certain conditions (Bogacz & Gurney, 2007). Although we have not simulated a task with progressive information accumulation, we believe our model is consistent with those findings and theories. We would map the cortical accumulation of evidence to the Proposer model component, and predict that the accumulation of activity in basal ganglia merely reflects that sensory activity, as the Actor component should simply approve each proposed action in a task with reliable reward history but difficult sensory judgments.

**Risky and Biased decision-making**

Our manipulations addressing risky decision-making are a modest first effort to apply our model and theory to addressing the individual differences, or risk factors, for risky decision-making. Our manipulation of D1/Go vs D2/NoGo pathway strength in dorsal striatum matched previous empirical results (Stopper, Khayambashi, & Floresco, 2013). We think these results suggest the utility of such a rich and detailed model in addressing the biological and environmental causes of risky behavior, but they certainly are not comprehensive. It remains for future work to expand on those predictions. Simulating a different task that better captures real-world risky decision-making would be useful in more fully capturing that phenomena. Such a task would be closer to a gambling task: payoffs are highly stochastic, with relatively little to learn about good and bad options.

The second key factor in human risky decision-making is capturing how humans learn about highly risky activities without directly experiencing the worst consequences. It seems clear that humans learn from vicarious experience; for instance, hearing about an auto accident caused by drunk driving seems to change behavior, despite a lack of firsthand experience with the outcome. Mental simulation has strong theoretical and empirical support (Reviewed in Barsalou, 2008). The precise mechanisms of vicarious learning are an important outstanding question, one we hope to address in future work. In particular, it is not known whether the dopamine system participates in vicarious learning, or whether that direct signal of reward and predicted reward is reserved for real rewards. Understanding how vicarious learning works in relation to risky decisions should have important implications for interventions, since most interventions involve communicating information about outcomes, rather than actual, experienced outcomes.

Risky decision-making has a good deal of overlap with bad, or biased, decision-making. Many paradigms do not distinguish risk-seeking behavior (in which some individuals prefer risk-reward tradeoffs in which high risks produce equally great average rewards) from bad or biased decision-making, in which some individuals simply make bad decision in certain domains, such as domains in which bad outcomes simply outweigh good ones, such as gambling against the house. Those decisions are also classified as examples of cognitive biases.

The current theory makes two important behavioral predictions regarding sources of biases. The first prediction is that some risky decision-making results from a failure to make a prediction-based decision. This strategy results from a tradeoff: the construction of useful predictive models is relatively time consuming, and so not worth performing for less important decisions. Such a time tradeoff has been proposed as one major component in mental effort effects (Shenhav, Musslick, Lieder, Kool, Griffiths, Cohen, & Botvinick, 2017; Kurzban, Duckworth, Kable, & Myers, 2013). Impulsive individuals, who are more vulnerable to making risky decisions (Białaszek, Gaik, McGoun, & Zielonka, 2015), may be biased in their preference for this time-saving tradeoff.

The second prediction is that, even when some amount of predictive model-building is performed, the model, and therefore the decision, will be biased by several factors. Most predictive models will by necessity be incomplete, since most domains do not allow for a construction of all paths to all outcomes in finite time (Daw & Dayan, 2014). The subset of outcomes that are predicted may be non-random and biased. One important bias should arise from motivated reasoning effects. For instance, if it is more pleasant to think about positive outcomes, people will include more pleasant than unpleasant outcomes in their model than an unbiased estimate. This bias should occur because our model posits that outcome prediction models are created based on further iterations of the same canonical circuit, other areas of PFC "decide" whether to create each predictive step in the model.

The behavior of the basal ganglia Actor component of each such predictive loop is shaped by and so ultimately under the control of dopamine reward signals. Because those signals sum total predicted reward across time and dimensions, (Schultz, 2013), this system does not produce locally optimal decisions. Of particular importance, decision-making in challenging domains appears to be highly biased toward perceived social reward, which is one type of *motivated reasoning* (Kahan, Jenkins-Smith, Braman, 2011). For instance, one may anticipate a proximal monetary reward for getting the right answer to a simple math problem, but also anticipate a social reward from peers for getting the answer that accords with their political beliefs (Kahen et al 2011; Kahan, Peters, Dawson, & Slovic, 2017). Motivated reasoning has been proposed as an underlying cause for flawed decisions of enormous consequence, such as the intelligence community deciding that Iraq likely possessed weapons of mass destruction (Jacobson, 2010). Understanding the neural basis of decision-making should help us understand and compensate for motivated reasoning, and perhaps other important biases. Our current model does model basal-ganglia based and dopamine-trained control components of the loops involved in creating predictions, so capturing such an effect remains for future work.

**Conclusions**

We have presented a relatively computationally and mechanistically detailed theory of how human beings make decisions. We presented a computational model that captures the core of that theory, as a canonical brain circuit. The mechanisms we proposed for this microcircuit are based on extensive empirical work on animal action selection, so the most central proposal is that human complex decision-making uses similar mechanisms and computations, enhanced by using more abstract representations, and more iterative and hierarchical steps. To allow such iteration to accumulate useful sub-decisions into a complex decision, each step must work serially by considering a single proposed action, plan, or conclusion at a time.

We term that canonical decision circuit a Proposer-Predictor-Actor-Critic model, in which the cortex proposes a potential plan or action, the basal ganglia acts to accept or reject that plan, and the amygdala and associated subcortical systems acts as a critic to gauge the success of that plan relative to expectations. This basic process can be enhanced to incorporate specific predictions of outcomes by the use of additional iterations of such a circuit to serve as a Predictor component, which can provide additional information to the Actor at the cost of extra time. When it uses this process of creating predictions, the circuit is performing model (or prediction)-based computations; when the Predictor component is not involved, the computations are largely model (or prediction)-free. (Although the learning in the other components may induce some neurons to represent likely outcomes, and so constitute a limited form of model-based processing).

While the empirical support for such a canonical circuit, and its use in human decision-making is indirect, we think it is quite strong. However, it remains for future work to investigate how such a circuit might work in detail, and whether and how a series of relatively simple choices can aggregate to create the most complex human planning, decision-making, and thinking.

**Acknowledgements**

This work was funded by NSF grant 5R01GM109996-03, "Integrated Cognitive Modelling for Risky Decision-Making"; Office of Naval Research contract D00014-12-C-0638, "Neural Mechanisms of Human Adaptive Executive Control", and ONR contract N00014-18-C-2067, "Capturing the Power and Pitfall of Human Decision-Making".

The authors thank Stephen Read and Thomas Hazy for extensive discussions of the theory, and Prescott Mackie for extensive programming and design of an early version of the model.